\documentclass[aps,pra,twocolumn,superscriptaddress]{revtex4-2}
\usepackage[utf8]{inputenc}
\usepackage{amsmath}
\usepackage{hyperref}
\usepackage{mathdots}
\usepackage{dsfont}
\usepackage{amsfonts}
\usepackage{amssymb}
\usepackage{graphicx}
\usepackage{filecontents}
\usepackage{tikz}
\usetikzlibrary{quantikz}
\usepackage{multirow}
\usepackage{xcolor}
\usepackage{bbold}

\begin{document}

\title{Entanglement Classification via Witness Operators generated by Support Vector Machine}

\author{Claudio Sanavio}
\email{claudio.sanavio@unibo.it}
\affiliation{Dipartimento di Fisica e Astronomia dell'Universit\`a di Bologna, I-40127 Bologna, Italy}
\affiliation{INFN, Sezione di Bologna, I-40127 Bologna, Italy}

\author{Edoardo Tignone}
\affiliation{Leithà S.r.l.~\text{\textbar} Unipol Group, Bologna, Italy}

\author{Elisa Ercolessi}
\affiliation{Dipartimento di Fisica e Astronomia dell'Universit\`a di Bologna, I-40127 Bologna, Italy}
\affiliation{INFN, Sezione di Bologna, I-40127 Bologna, Italy}

\begin{abstract}
Although entanglement is a basic resource for reaching quantum advantange in many computation and information  protocols, we lack a universal recipe for detecting it, 
with analytical results obtained for low dimensional systems and few special cases of higher dimensional systems. 
In this work, we use a machine learning algorithm, the support vector machine with polynomial kernel, to classify separable and entangled states. 
We apply it to two-qubit and three-qubit systems, and we show that, after training, the support vector machine is able to recognize if a random state is entangled with an accuracy up to $92\%$ for the two-qubit system and up to $98\%$ for the three-qubit system. 
We also describe why and in what regime the support vector machine algorithm is able to implement the evaluation of an entanglement witness operator applied to many copies of the state, and we describe how we can translate this procedure into a quantum circuit.
\end{abstract}

\maketitle

\section{Introduction}\label{sec:I}

Quantum entanglement~\cite{Horodecki2009} is one of the main features that distinguish between classical and quantum states and represents one of the basic ingredients for reaching quantum advantage in computation and information protocols~\cite{Jozsa2003,Wootters1998a,Guhne2009}. 
An $N$-particle state is entangled if its density operator can not be written as tensor product or sum of tensor products of $N$ single-particle density operators or, in other words, if 
the state of the system cannot be entirely described considering its $N$ single components only. Despite the simplicity of such a definition, the problem of identifying, classifying and quantifying entanglement is mathematically extremely hard. Many analytical and numerical results have been obtained in the study of bipartite entanglement, including the well known positive partial transpose (PPT) criterium, which establishes that, for two-qubit and qubit-qutrit states, the positivity of the partial transposition of the density operator provides a necessary~\cite{Peres1996} and sufficient~\cite{Horodecki1996} condition for entanglement.
In spaces of higher dimension, however, the PPT criterion offers only a necessary condition for separability ~\cite{Peres1996}, reflecting the fact that entanglement of an $N$-particle system (with $N\geq 3$) is much richer in comparison to the bipartite case. This is because quantum entanglement is a ``monogamous" property that can be shared among the different parts in many non-equivalent ways and not freely, since the degree of entanglement between any two of its parts influences the degree of entanglement that can be shared with a third part~\cite{Coffman2000,Osborne2006}. 
In the last two decades, there has been a great effort to recognize if a state with $N\geq3$ is entangled and many analytic results have been found, see the review papers~\cite{Eltschka2014,Szalay2015}. However, even in the simplest case of three qubits, a general solution is still missing.

In this paper we tackle the still open question of finding if a given state is entangled or separable as a classification problem. \\
Many classification problems can be solved efficiently by machine learning (ML) algorithms~\cite{Kotsiantis2006,Scholkopf2002}, whose goal is to find a way to infer, from the available labeled data, the class of an unlabeled data point. Neural network ML algorithms for classification of entangled states have already been adopted in previous works, such as in Ref.~\cite{Ma2018} for the classification of two-qubit states labeled by means of Bell’s inequalities, or in Ref.~\cite{Harney2020}, where a classifier is trained to recognize separable states of multi-qubit systems.\\
Another well-known ML algorithm is the Support Vector Machine (SVM)~\cite{BoserVapnik1992}, that has been proved to be very solid in solving classification problems in image recognition~\cite{Chandra2021}, medicine~\cite{Hazra2016,Kadam2020} and biophysics~\cite{Cai2001} among others. A collection of applications of the SVM algorithm can be found in the review paper~\cite{Cervantes2020}. The SVM approach has been recently investigated for classification of two-qubit states~\cite{Lu2018} and four-qubit states~\cite{Vintskevich2022}. In particular, in the former paper,  the authors use a SVM with exponential kernel, supported by a preliminary Convex Hull Approximation (CHA) analysis, to construct a separability-entanglement classifier via a supervised learning approach. 

In this work, we put forward a new approach, exploiting the so-called kernel trick~\cite{Scholkopf2002} with polynomial kernels,  in which we train a SVM to classify entangled and separable states with a high predictive power, specilizing the analysis first to the case of the two-qubit system, as a benchmark, and then generalizing it to the classification of different classes of entangled states of a three-qubit system. 
The performance of the algorithm is evaluated through the following metrics: the accuracy $a$, corresponding to the percentage of true positives, i.e.~the (either separable or entangled) states that are correctly classified, the precision $p$, that measures the percentage of predicted entangled states that are really entangled, and the recall $r$, that measures the percentage of entangled states that are correctly classified. The hyperparameters of the algorithm are chosen to get $p=1$ and the maximum possible value of $r$ on the validation set.\\
In this way we are effectively implementing an {\it entanglement witness} operator, which does not misclassifies separable states as entangled while also minimizing the number of entangled states that are misclassified as separable. Such witness operator is explicitly constructed for a linear kernel, obtaining a standard witness functional over the space of density matrices of the system, as well as in the case of a non-linear polynomial kernel of degree $n$. In the latter case, which corresponds to the construction of a witness functional over $n$-copies of the system, the performance of the algorithm increases. Finally, we present a digital quantum circuit which is able to implement the construction and the evaluation of such entanglement witnesses. 

The content of the paper is as follows. In Section~\ref{sec:II} we review the entanglement classification for two-qubit and three-qubit systems, with emphasis on the analytical results that we use to generate the dataset of pure and mixed states~\cite{Eltschka2014,Dur2000,Eltschka2008,Eltschka2012,Jung2009,Hashemi2012}. In Section~\ref{sec:III} we describe the SVM algorithm and the kernel trick. In Section~\ref{sec:IV} we relate the separating hypersurface obtained via the SVM to the evaluation of an entanglement witness operator, providing for that also an operational procedure that can be implemented on a digital quantum computer.

In Section~\ref{sec:V} we train the SVM with two-qubit states for which we have an exact classification, i.e. the PPT criterion, and eventually test our method on a large sample. Section \ref{sec:VI} is devoted to the extension of our  analysis to three-qubit systems. Finally, in Section~\ref{sec:VII} we draw our conclusion and give outlooks for possible future research directions. 

\section{Entanglement classification}\label{sec:II}

In this section we give a brief overview of the entanglement in two-qubit and three-qubit systems, with emphasis on the different analytical results found in the literature to classify entangled and separable states. 

Let us first consider a pure quantum state $|\psi\rangle$ defined in the Hilbert space $\mathcal{H}^{\otimes N}$ of $N$ identical particles and 
described by the density operator $\hat{\rho}=|\psi\rangle\langle\psi|$.

The state $\hat{\rho}$ can be written as a product state of $M\leq N$ density operators~\cite{NielsenChuang2000} as
\begin{eqnarray}\label{eq:separablestate}
\hat{\rho}=\otimes_{j=1}^{M}\hat{\rho}_{j}
\end{eqnarray} 

\noindent  If $M=N$, the state is called fully separable, and each $\hat{\rho}_j$ acts on the one-particle space $\mathcal{H}$. If $M<N$, at least one of the density operators $\hat{\rho}_j$ describes more than one particle. If $M=1$, the state $\hat{\rho}$ is called Genuinely Multipartite Entangled (GME).
  
In the case of a two-qubit state, with $N=2$ and $\mathcal{H}=\mathbb{C}^2$, the system can either be separable, with $M=2$,  or entangled, with $M=1$. 
In case the system is composed by three qubits,
labeled as A,B and C, we have $N=3$ and $\mathcal{H}=\mathbb{C}^2$. The state is fully separable (SEP) when $M=3$, and therefore Eq.\eqref{eq:separablestate} reads as $\hat{\rho}=\hat{\rho}_A\otimes\hat{\rho}_B\otimes\hat{\rho}_C$. When $M=2$ the state is called biseparable, and the density operator can be written as $\hat{\rho}_A\otimes\hat{\rho}_{BC}$ (A-BC), $\hat{\rho}_B\otimes\hat{\rho}_{AC}$ (B-AC) or $\hat{\rho}_C\otimes\hat{\rho}_{AB}$ (C-AB). When $M=1$ the state is GME. GME states play an important role to devise robust distribution protocols~\cite{Apollaro2020}.

\noindent Mixed states are a statistical mixture of pure states~\cite{Horodecki2009}, and are described by a density operator of the form
\begin{equation}\label{eq:mixture}
\hat{\rho}=\sum_ip_i\hat{\rho_i},
\end{equation}
where $\sum_ip_i=1$ and $\hat{\rho}_i$ are pure states.
The mixed state $\hat{\rho}$ is fully separable when each pure state $\hat{\rho}_i$ is fully separable.

\noindent For a two-qubit system (as well as for a qubit-qutrit system), the PPT criterion \cite{Peres1996,Horodecki1996}
offers a necessary and sufficient condition for labeling a quantum state as entangled: a two-qubit state is separable if and only if the partial transposition of the density matrix returns a density matrix. In all other cases, the PPT criterion offers only a necessary condition for separability ~\cite{Peres1996}.

\begin{figure}[ht!]
\centering
\includegraphics[scale=0.38]{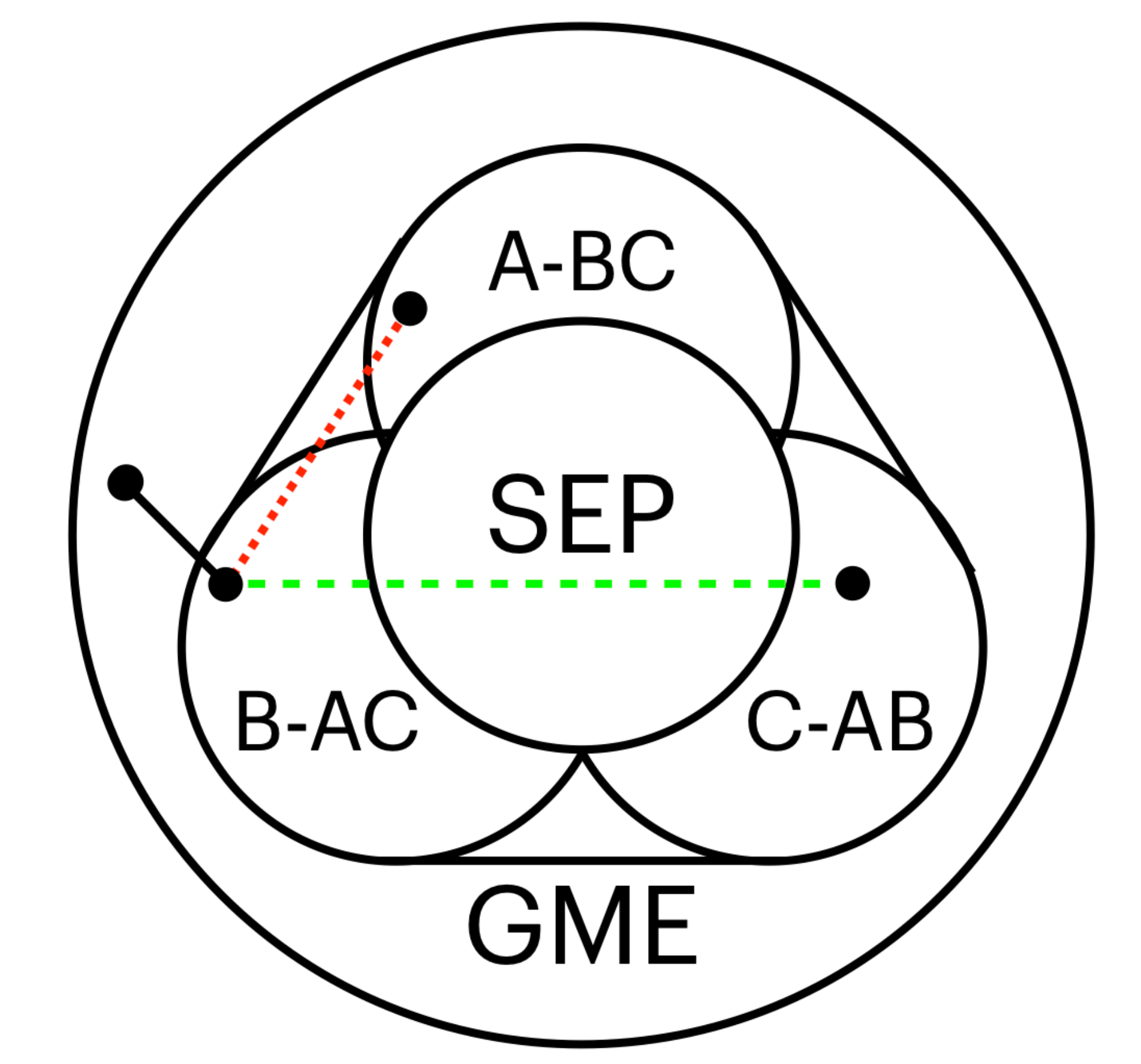}
\caption{The set of three-qubit states, called A, B and C, in terms of their level of entanglement. GME is the set of geuinely multipartite entangled states, SEP is the set of fully separable states, and A-BC, B-AC, C-AB represent the three possible factorization of the density operator in the biseparable case. Lines between sets of biseparable states represent the convex hull of biseparable states.
The states are represented by circles. The segments joining them are their convex combinations. In the figure are represented convex combination of a state in B-AC with states in GME (black continuous line), in C-AB (green dashed line), and in A-BC (red dotted line). Depending on the weights given to the convex combination, the resulting state can lie in a different subset.\label{fig:setthreequbit}}
\end{figure}

\noindent When considering the statistical mixture of states of three qubits, it is not straightforward to tell if the state is separable, biseparable or GME. Fig.~\ref{fig:setthreequbit} shows the general classification of three-qubits states. The convex set of fully separable states is represented by the innermost circle, which is sorrounded by the three kinds of biseparable states (A-BC,B-AC,C-AB) and finally by the GME states (largest circle). The segment joining two points represents the convex combination of two states. A statistical mixture of states can lie in any of the aforementioned classes. When the state is a statistical mixture of biseparable states, with respect to different partitions, the result can either be fully separable, or biseparable with respect to one of the partitions (green dashed segment in Fig.\ref{fig:setthreequbit}), or biseparable with respect to none of the partitions (red dotted segment in Fig.\ref{fig:setthreequbit}), thus forming the convex-hull of the biseparable states. It is known that all GME states can be obtained by applying SLOCC (stochastic local operations and classical communication) operations~\cite{Dur2000} either on the GHZ state
\begin{eqnarray}\label{eq:GHZ3}
|\text{GHZ}\rangle &=& \frac{1}{\sqrt{2}}(|000\rangle+|111\rangle),
\end{eqnarray}
or on the W state
\begin{eqnarray}\label{eq:W3}
|\text{W}\rangle &=& \frac{1}{\sqrt{3}}(|100\rangle+|010\rangle+|001\rangle).
\end{eqnarray}

To correctly label as entangled or separable the three-qubit states used to construct the dataset for the ML classifier,
we will resort to the analytical calculation of several entanglement measures~\cite{Plenio2007}. In the remaining of this section we are going to introduce these analytical results, and we refer to Appendix~\ref{app:I} for the description of the states. In general, an entanglement measure $\mu$ is a real function defined on the set of the states that does not increase under the application of a SLOCC transformation~\cite{Dur2000}, for which two states belong to the same class if you can transform one into the other with an invertible local operation, and the application of SLOCC brings a state into a class with lower or equal value of any entanglement measure. 

\noindent Pure states can be uniquely~\cite{Dur2000} labeled with the entanglement entropy $S$, the GME-concurrence $C_{\text{GME}}$ and the three-tangle $\tau$.
The entanglement entropy $S_m$ ~\cite{Vedral2002} corresponds to the Von-Neumann entropy calculated on the reduced system $\hat{\rho}_{m}=\text{Tr}_{ABC-m}[\hat{\rho}]$, with $m=$A,B,C: 
\begin{equation}\label{eq:EntanglementEntropy}
S_m=-\text{Tr}[\hat{\rho}_m\ln\hat{\rho}_m].
\end{equation}
Genuinely Multipartite Entanglement concurrence~\cite{Wootters1998b} is defined as
\begin{equation}\label{eq:GMEconcurrence}
C_{\text{GME}}=2\min_{i\in \mathcal{P}}\sqrt{1-\text{Tr}[\hat{\rho}_{i}^2]},
\end{equation}
where $\mathcal{P}=\{AB,BC,AC\}$ is the set of the possible sub-systems. Finally, the three-tangle $\tau$~\cite{Coffman2000} accounts for the genuinely multipartite entanglement of the GHZ class.
Given the reduced density operator $\hat{\rho}_{AB}$ of the subsystem of the qubits A and B, we define the matrix $\tilde{\rho}_{AB} = (\sigma_y\otimes\sigma_y)\hat{\rho}_{AB}(\sigma_y\otimes\sigma_y)$, and we denote assam $\lambda_{AB}^{1},\lambda_{AB}^{2}$ the square root of the first and second eigenvalues of $\tilde{\rho}_{AB}\hat{\rho}_{AB}$. Then, the three-tangle is
\begin{equation}\label{eq:threetangle}
\tau(\hat{\rho})=2(\lambda_{AB}^1\lambda_{AB}^2+\lambda_{AC}^1\lambda_{AC}^2),
\end{equation}
with $\lambda^{i}_{AC}$ defined in a similar manner as $\lambda^{i}_{AB}$, for $i=1,2$. 

\noindent In the case of mixed states, the entanglement measures defined above fail to classify entangled states. One needs to calculate the Convex-Roof Extension (CRE)~\cite{Uhlmann1998} of an entanglement measure $\mu$ defined as
\begin{equation}
\tilde{\mu}(\hat{\rho})=\inf_{P}\sum_i p_i\mu(\hat{\rho}_i),
\end{equation}
where $P$ stands for all the possible decompositions $\{p_i,\hat{\rho}_i\}$ such that $\hat{\rho}=\sum_ip_i\hat{\rho}_i$, with $\sum_ip_i=1$ and $\hat{\rho}_i$ is a pure state.
This is in general difficult to calculate, and numerical solutions are usually very costly. 
Therefore, we use particular types of mixed states for which the analytic solutions for the CRE of the three-tangle~\cite{Jung2009,Hashemi2012,Lohmayer2006}, and the CRE of GME-concurrence~\cite{Eltschka2014,Eltschka2008,Eltschka2012} are available. These are the GHZ-symmetric states, the X-states, and the statistical mixture of GHZ and W states which are defined in Appendix~\ref{app:I}, which will be used in the following to construct the three-qubit dataset of  the SVM algorithm.
Table \ref{tab:tripartiteclassification} summarizes how the labels are assigned to these different type of states.
\begin{table}
\begin{center}
\begin{tabular}{| c | c | c | c |} 
 \hline
 \multirow{2}{*}{Class} & \multirow{2}{*}{Entanglement measure} & \multicolumn{2}{c|}{label}\\ [0.5ex]
 & & SEP-vs-all & GME-vs-all \\
 \hline
 \hline
 SEP & $S_m=0$ for $m=A,B,C$ & -1 & -1 \\
 \hline 
 A-BC &  $S_A=0$ and $S_B,S_C>0$  & +1 & -1\\
 B-AC &  $S_B=0$ and $S_A,S_C>0$  & +1 & -1\\
 C-AB &  $S_C=0$ and $S_A,S_B>0$  & +1 & -1\\
 \hline
 GHZ & $S_A,S_B,S_C>0$ and $\tau>0$  & +1 & +1\\
 X-state & $\tilde{C}_{\text{GME}}>0$& +1 & +1\\
 GHZ-sym & $\tilde{C}_{\text{GME}}>0$ and $\tilde{\tau}>0$& +1 & +1\\
 GHZ+W & $\tilde{\tau}>0$ & +1 & +1\\
\hline
\end{tabular}
\end{center}
\caption{The classification of three-qubit states into entanglement classes. SEP are pure and mixed separable states. A-BC, B-AC, C-AB are the biseparable states. GME states divide into pure GHZ states, X-state, GHZ symmetric and statistical mixture of GHZ and W states, each labeled accordingly to the corresponding entanglement measures. The mixed states are classified using the CRE of the entanglement measures, denoted as  $\tilde{C}_{\text{GME}}$ and $\tilde{\tau}$. Fully separable states, pure and mixed, are labeled as $-1$ for both the SEP-vs-all and GME-vs-all classifiers, biseparable states are labeled as $+1$ in the SEP-vs-all classifier and $-1$ in the GME-vs-all classifier. GME states are always labeled as $+1$.\label{tab:tripartiteclassification}}
\end{table}

\section{Support Vector Machine}\label{sec:III}

In this section we give a general introduction of the SVM.

Suppose that we have a dataset divided into two classes, composed of $m$ pairs $\{(\mathbf{x}_i,y_i)\}$. The $i$-th observation $\mathbf{x}_i\in\mathbb{R}^d$ describes the characteristics of the point while $y_i=\pm1$ labels its class. The SVM~\cite{BoserVapnik1992,Scholkopf2002} is a ML classification algorithm based on the assumption that a hyperplane, or more generally a $d-1$ surface, which separates the two classes in the feature space exists. The hyperplane is defined through a linear decision function $f(\mathbf{x})$, with parameters $\mathbf{w}\in\mathbb{R}^d$, called weights, and $b\in\mathbb{R}$, called bias, given as
\begin{eqnarray}\label{eq:decisionfunction}
f(\mathbf{x})&=& \mathbf{w}^T\mathbf{x}+b,\\
\text{s.t.}&&
\begin{cases}
f(\mathbf{x}_i)>0\quad\text{if}\quad y_i=+1,\\
f(\mathbf{x}_i)<0\quad\text{if}\quad y_i=-1.
\end{cases}\nonumber
\end{eqnarray}

\begin{figure}[ht!]
\centering
\includegraphics[scale = 0.28]{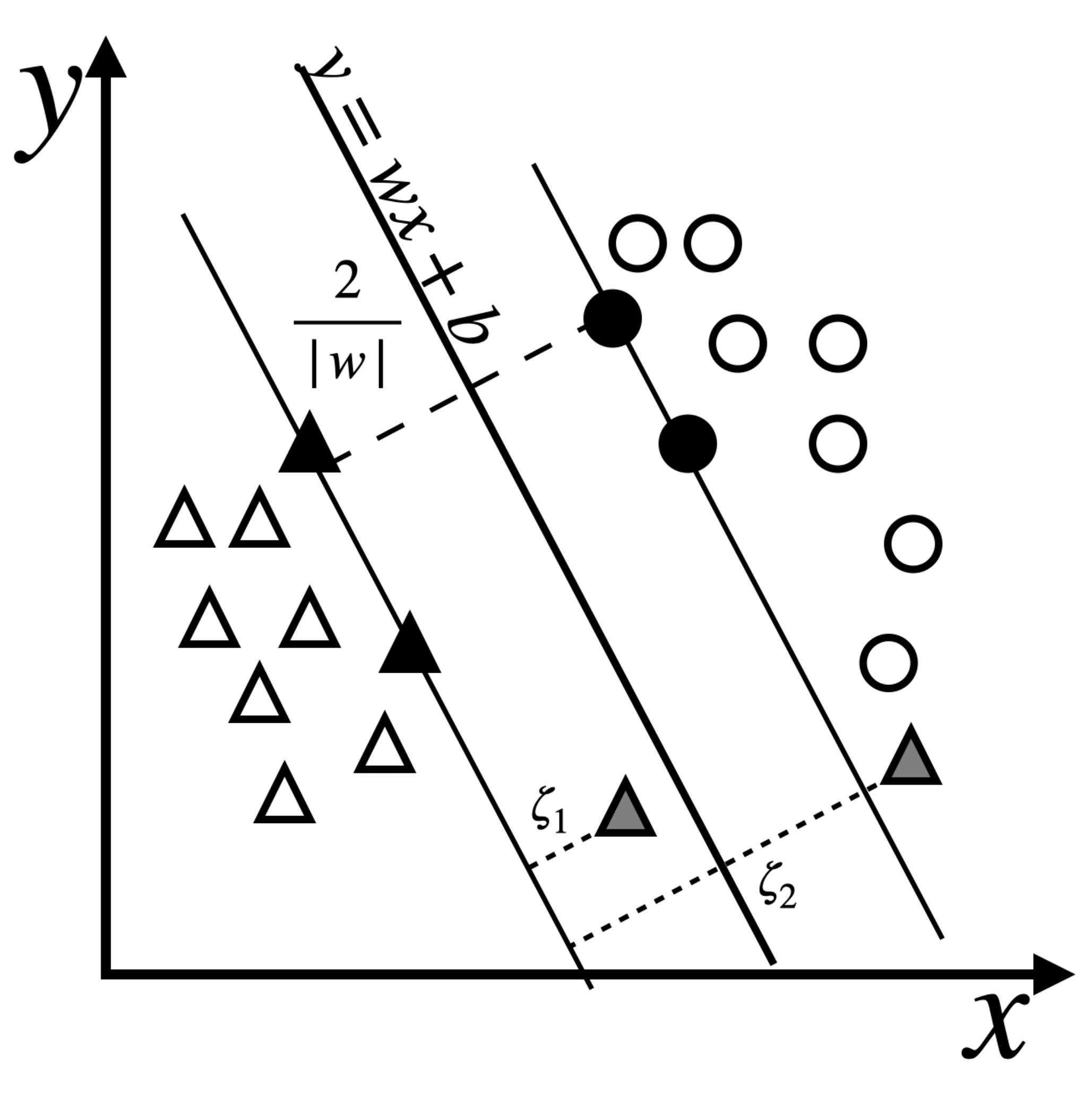}
\caption{Representation of the separating line between two data classes, the circles and the triangles, in a two-dimensional plane. The two classes are linearly separable by the straight line $y=wx+b$. The margin, i.e.~the distance between the hyperplane and the closest data point of each class is $1/|w|$. The lines $y-wx-b=\pm1$ cross the support vectors, here drawn with a filled shape. The slack variables $\zeta_1$ and $\zeta_2$ penalize the possible misclassification of the two gray triangle points. \label{fig:SVMgeometry}}
\end{figure}

\noindent The SVM algorithm aims to find the hyperplane $(\mathbf{w},b)$ that maximizes the margin $1/\lVert \mathbf{w}\rVert$ ($\lVert \cdot\rVert$ being the euclidean norm) 
 between itself and the closest points of the two classes. Fig.~\ref{fig:SVMgeometry} shows a two-dimensional dataset made of two classes, the circles and the trangles, that are separated by a hyperplane.

\noindent Maximizing the margin corresponds to finding 
\begin{eqnarray}\label{eq:maxmargin}
&&\min_{\mathbf{w}}\frac{1}{2}\lVert \mathbf{w}\rVert^2,\\
&&\text{s.t.}\quad
y_i(\mathbf{w}^T\mathbf{x}_i+b)\geq 1,\text{ for }i=1,\dots,m.
\nonumber
\end{eqnarray}
We can assume that the hyperplane defined by the decision function $f$ is not able to perfectly separate the two classes. Thus, we can introduce $m$ slack variables $\zeta_i$, one for each observation $\mathbf{x}_i$, to account for the constraints of $f$ in Eq.~\eqref{eq:decisionfunction} that are not satisifed. Hence, Eq.\eqref{eq:maxmargin} is refined into

\begin{subequations}
\begin{eqnarray}\label{eq:SVMslack}
&\min_{\mathbf{w},\zeta}&\frac{1}{2}||\mathbf{w}||^2+\sum_{i=1}^m\lambda_{y_i}\zeta_i,\\
&\text{s.t.}&y_i(\mathbf{w}^T \mathbf{x}_i+b)\geq 1-\zeta_i,
\label{eq:SVMslackconstrainta}\\
&&\zeta_i\geq0.
\label{eq:SVMslackconstraintb}
\end{eqnarray}
\end{subequations}
The constants $\lambda_{y_i}\geq 0$ are called regularization parameters. Geometrically, the action of the slack variables $\zeta_i$ can be seen as a non-smooth deviation of the hyperplane, as shown in Fig. \ref{fig:SVMgeometry}.

In order to construct a Lagrange function from which we can variationally derive Eq.~\eqref{eq:SVMslack}, we use the so called Karush-Kuhn-Tucker (KKT) complementary conditions~\cite{Scholkopf2002}: after introducing dual variables $\alpha_i$ and $\lambda_{y_i}$, the constraints can be re-expressed as: 
\begin{subequations}
\begin{eqnarray}\label{eq:KKTrelation1}
&&\alpha_i[y_i(\mathbf{w}_i^T\mathbf{x}_i+b)-1+\zeta_i]=0, \\
&&\zeta_i(\alpha_i-\lambda_{y_i})=0. \label{eq:KKTrelation2}
\end{eqnarray}
\end{subequations}

\noindent These KKT conditions state that if the variable $\mathbf{x}_i$ lies on the correct side of the hyperplane, with $y_i(\mathbf{w}_i^T\mathbf{x}_i+b)>1$, then $\alpha_i=0$ and $\zeta_i=0$. On the other hand, if the variable $\mathbf{x}_i$ lies exactly on the margin, with $y_i(\mathbf{w}_i^T\mathbf{x}_i+b)=1$, then $\alpha_i =\lambda_{y_i}$, and $\zeta_i=0$. Lastly, if the point $\mathbf{x}_i$ lies on the wrong side of the hyperplane, we have $0<\alpha_i<\lambda_{y_i}$ and $\zeta_i>0$. In general, a number $m'<m$ of points $\mathbf{x}_i$ have a corresponding dual variable $\alpha_i\neq0$. These points, that lie on the separating hypersurface, are called support vectors.

\noindent The KKT conditions allow us to write the Lagrange function of the problem as 
\begin{eqnarray}\label{eq:primalproblem}
\mathcal{L} &=& \frac{1}{2}\lVert \mathbf{w}\rVert^2+ \sum_{i=1}^m \alpha_i[1-\zeta_i-y_i(\mathbf{w}^T\mathbf{x}_i+b)]\nonumber\\
&&+ \sum_{i=1}^m   \lambda_{y_i}\zeta_i- \sum_{i=1}^m \beta_i\zeta_i,
\end{eqnarray}
where $\alpha_i,\beta_i\geq0$ are the Lagrange multipliers introduced for the constraints
~\eqref{eq:SVMslackconstrainta} 
and \eqref{eq:SVMslackconstraintb} respectively. The optimality condition is satisfied when the partial derivatives with respect to the primal variables $\mathbf{w},b,\zeta_i$ vanish,
\begin{subequations}
\begin{eqnarray}
\partial_{\mathbf{w}}\mathcal{L} & = & \mathbf{w}-\sum_{i=1}^m y_i\alpha_i\mathbf{x}_i=0\label{eq:La}\\
\partial_b\mathcal{L} & = & -\sum_{i=1}^my_i\alpha_i=0\label{eq:Lb}\\
\partial_{\zeta_i}\mathcal{L} & = & \lambda_{y_i}-\alpha_i-\beta_i=0, \forall i.\label{eq:Lc}
\end{eqnarray}
\end{subequations}
Substituting Eqs.~\eqref{eq:La},\eqref{eq:Lb} and \eqref{eq:Lc} into Eq.~\eqref{eq:primalproblem} yields the Lagrange function expressed in terms of the multipliers $\mathbb{\alpha}=(\alpha_1,\dots,\alpha_m)$
\begin{eqnarray}\label{eq:Lagrangesolution}
\mathcal{L}(\mathbb{\alpha})&=&\sum_{i=1}^m\alpha_i-\frac{1}{2}\sum_{i,j=1}^{m}\alpha_iy_iK_{ij}y_j\alpha_j,
\end{eqnarray}
where $K$ is called kernel matrix, and the component $K_{ij}$ corresponds to the inner product $\langle \mathbf{x}_i,\mathbf{x}_j\rangle$ between the feature vectors.

The matrix $K$ encodes the power of the SVM algorithm, that is able to perform a separation with a higher-degree decision function. In principle, to do so one has to embed the feature vectors into a higher dimensional space with an embedding function $\Phi:\mathbb{R}^d\to\mathbb{R}^D$, with $D>d$, and find the linear hyperplane that separates the data classes in the embedding space $\mathbb{R}^D$. As a result, the Lagrange function can be written as in Eq.\eqref{eq:Lagrangesolution}, with $K_{ij}=\langle \Phi(\mathbf{x}_i),\Phi(\mathbf{x}_j)\rangle$ being the inner product in the embedding space.
The procedure of adopting a specific kernel matrix $K$ in order to move to a higher-dimensional space is known as kernel trick~\cite{Scholkopf2002}. \\
One common choice is the polynomial kernel of degree $n$, with $K(\mathbf{x},\mathbf{x}')=(\langle\mathbf{x},\mathbf{x}'\rangle+1)^n$. It corresponds to embedding the $d$ dimensional vectors $\mathbf{x}$ and $\mathbf{x}'$ into a space of dimension~\cite{Scholkopf2002}
\begin{equation}
D=\begin{pmatrix}
n+d-1\\
n
\end{pmatrix}
\end{equation}
where we can look for a separating $(D-1)$-surface. 
The components of $\Phi(\mathbf{x})$ in the $D$-dimensional space come from the $D$ terms in the multinomial expansion of degree $n$. 
The constant $1$ plays an important role, as it allows the presence of all the terms up to degree $n$ in the decision function $f$, while in its absence the decision function has only the terms of degree exactly $n$. As the value of $D$ is extremely large for real case scenarios, the kernel trick saves us from having to explicitly embed the data in a much larger space and offers a smart way of coping with the embedding process. 

The classification of a new data point $\mathbf{x}$ consists simply in calculating the decision function
\begin{equation}\label{eq:decisionfunction2}
f(\mathbf{x})=\sum_{i=1}^{m'}\alpha_i y_i K(\mathbf{x}_i,\mathbf{x})+b,
\end{equation}
and the label will be given by the sign of $f(x)$. Note that the sum is made only over the $m'$ support vectors, as for the other points the Lagrange multipliers are zero.

\section{SVM-derived entanglement witnesses}\label{sec:IV}

In this section we show how the separating hyperplane provided by the SVM algorithm as a result of the minimization in Eq.~\eqref{eq:SVMslack} can be translated into a Hermitian operator that we can measure to detect entanglement. We also describe a procedure that can be implemented on a quantum computer in order to measure its mean value. The hyperplane inferred by the SVM has a huge similarity with the concept of entanglement witness. 
The classification in the SVM is calculated by considering the sign of the decision function $f(x)$. In a similar manner, an entanglement witness operator $\hat{W}$ is such that if $\text{Tr}[\hat{W}\hat{\rho}]>0$ then the state $\hat{\rho}$ is entangled~\cite{NielsenChuang2000}, providing a sufficient but not necessary condition for entanglement detection. The convexity of separable states ensures us that we can find such operator~\cite{Horodecki2009}. 

\begin{figure}
\centering
\includegraphics[scale=0.13]{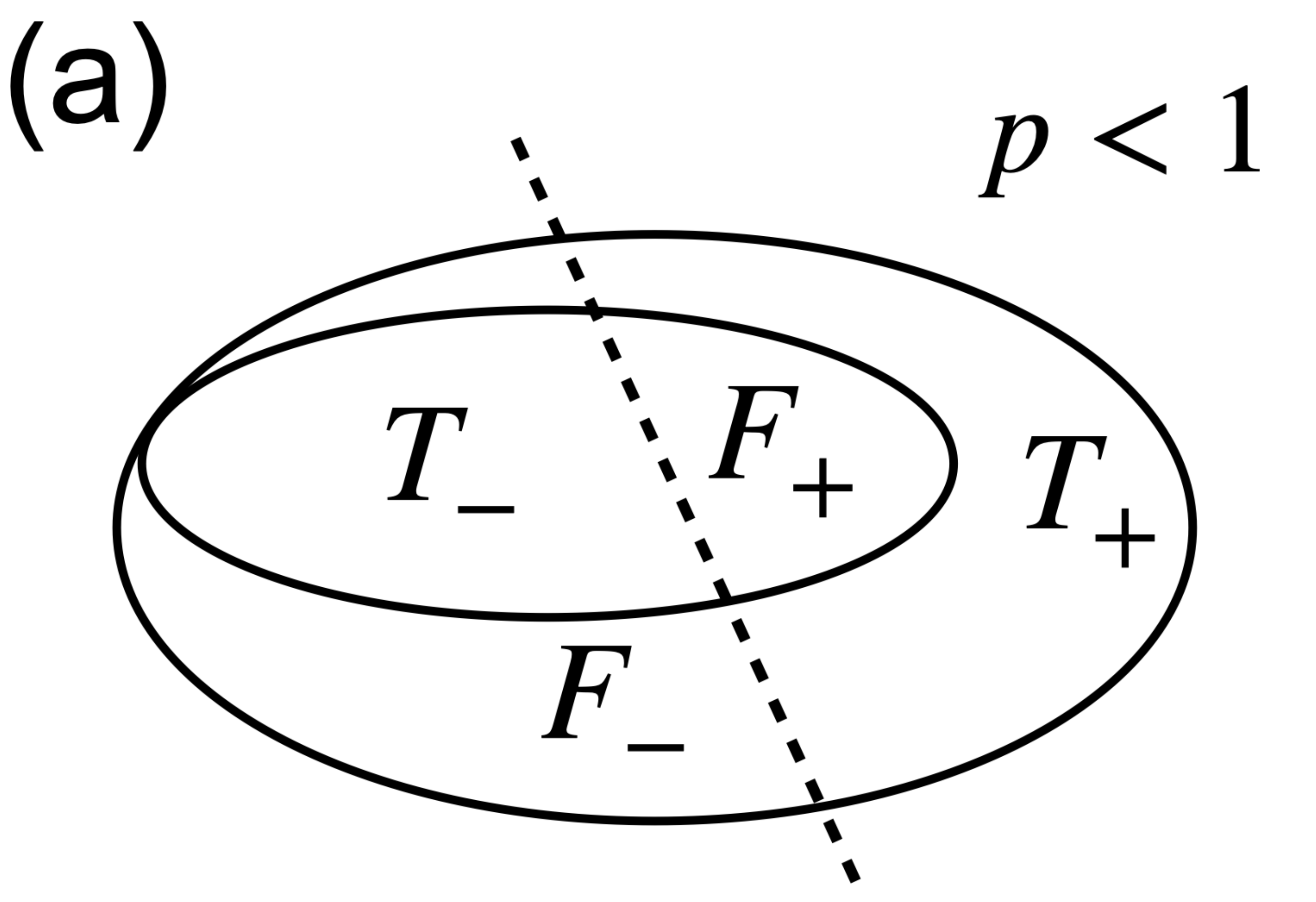}
\includegraphics[scale=0.13]{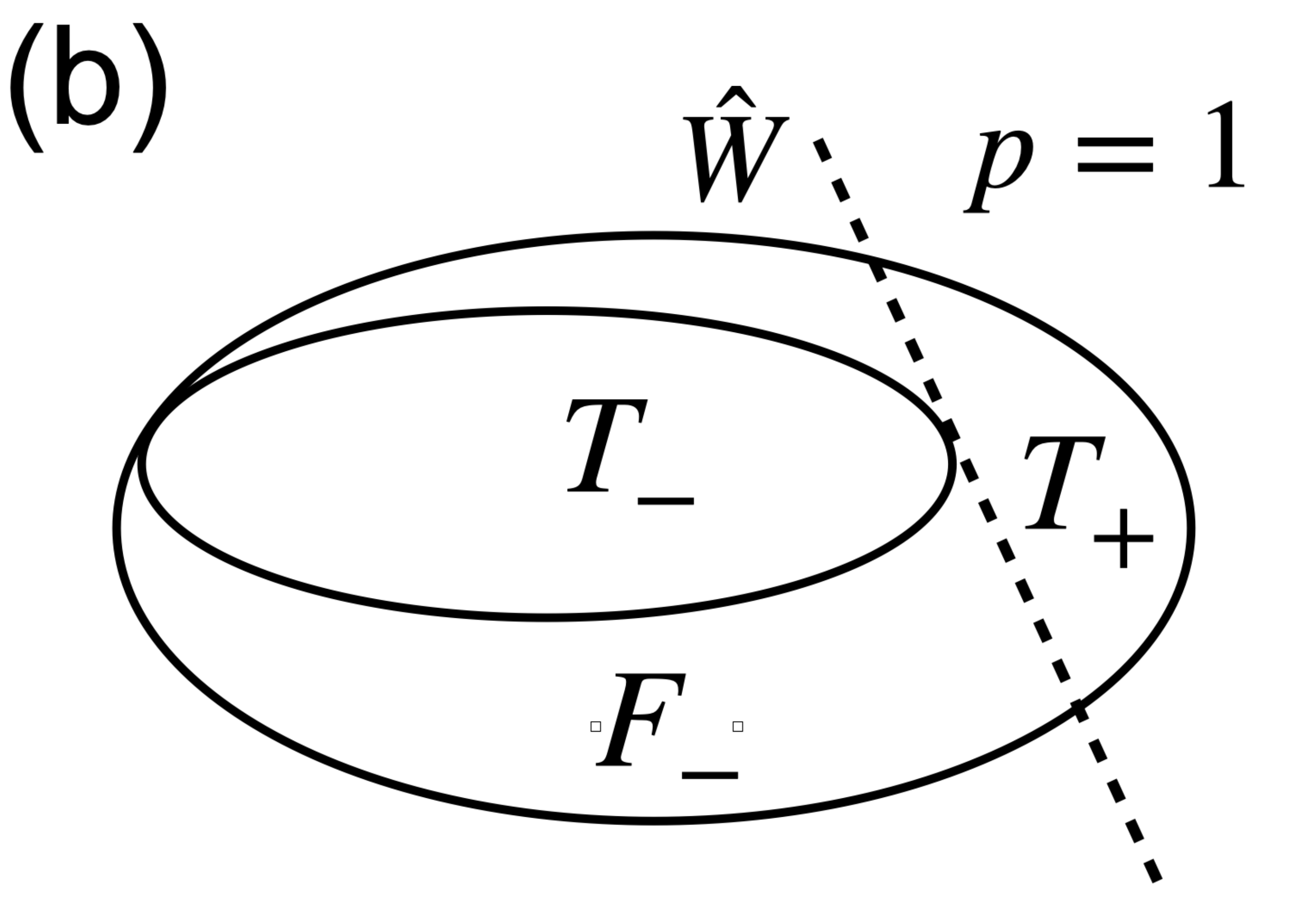}
\caption{Schematic representation of a
classification performed on a convex set. (a) the separating hyperplane, represented by the dotted line, divides the two classes $+1$ on the right, and $-1$ on the left, with a certain amount of true and false positives, $T_+,F_+$, and true and false negatives $T_{-},F_{-}$, therefore the precision $p<1$. (b) the separating hyperplane has been built to mimic the behavior of an entanglement witness operator $\hat{W}$. In this case, there are not false positives, and the precision $p=1.$\label{fig:recall}}
\end{figure}

Consistently with the choices of Table I, we assign to the separable states the label $-1$ and to the entangled states the label $+1$. In order to evaluate the performance of the classifier, we use the accuracy
\begin{equation}\label{eq:accuracy}
a=(T_{+}+T_{-})/m,
\end{equation}
where $m$ is the total number of states considered in the dataset. Here $T_{+}, T_{-}$ are the number of true positives/negatives repsectively, i.e.~the number of entangled/separable  states correctly classified as entangled/separable. Furthermore, the precision $p$ and the recall $r$ associated to the classification of entangled states are given by
\begin{eqnarray}
p&=&T_{+}/(T_{+}+F_{+}),\label{eq:precision}\\
r&=&T_{+}/(T_{+}+F_{-}),\label{eq:recall}
\end{eqnarray}
where $F_{+}, F_{-}$ represent the number of false positives/negatives respectively, i.e.~the number of
separable/entangled states misclassified as entangled/separable.
The condition $p=1$ indicates that we are not misclassifying separable states as entangled. Therefore, the hyperplane $(\mathbf{w},b)$ that we find with this condition respects the definition of entanglement witness, see Fig.~\ref{fig:recall}. We can get this condition by fine-tuning the regularization hyperparameters $\lambda_{+}$ and $\lambda_{-}$ defined in equation \eqref{eq:primalproblem}. In fact, we assign to each data point the regularization parameter $\lambda_+$ or $\lambda_{-}$ depending on whether it belongs to the class of entangled or separable states, respectively. 
The condition $\lambda_{+}>\lambda_{-}$  penalizes the misclassification of entangled states. On the other hand, the condition $\lambda_{-}>\lambda_{+}$ penalizes the misclassification of separable states, favoring thus the classifier to correctly predict $-1$, leading in turn to a higher precision in classification of entangled states at the cost of a lower recall. In terms of the classification algorithm, the witness operator $\hat{W}$ is related to a hyperplane with perfect precision of entangled states. 
When the number of false positive $F_{+}$ is minimized, we mimic an entanglement witness operator $\hat{W}$, with the property that $\text{Tr}[\hat{W}\hat{\rho}]>-b$ if and only if the state is entangled, with $b$ being the bias of the hyperplane, see Eq.~\eqref{eq:decisionfunction}.

The operation $\text{Tr}[\hat{W}\hat{\rho}]$ is related to a SVM hyperplane with linear kernel, $K(\mathbf{x}_i,\mathbf{x}_j)=\mathbf{x}_i^T\mathbf{x}_j$.
In the following, we will show that a more precise classification of entangled states can be carried out by non-linear kernels of degree $n$. The latter correspond to exotic witness operators $\hat{W}_n$ that work on $n$ copies of the state $\hat{\rho}$.

\noindent With little algebra one can show that the decision function \eqref{eq:decisionfunction2} can be written as 
\begin{equation}\label{eq:witnessent}
f_n(\hat{\rho})=\text{Tr}[\hat{W}_n\hat{\rho}^{\otimes n}]+b
\end{equation}
with the non-linear witness operator $\hat{W}_n$ given by
\begin{eqnarray}\label{eq:witnessSVM}
\hat{W}_n &=&\sum_{i=1}^{m'}\sum_{l=1}^{n}y_i\alpha_{i,l}(\hat{\omega}_i^{\otimes l}\otimes\mathds{1}_2^{(n-l)})
\end{eqnarray}
where $m'$ is the number of support vectors states. Here, $\mathds{1}_2^{(k)}$ indicates the identity matrix in the space of $k$ qubits, with dimension $2^k$. The coefficients of the operator $\hat{\omega}_i$  depend only on the $i-th$ support vector state, which can be found, together with the values of $\alpha_{i,l}$,  via a classical training of the SVM classifier.

In this way, the classification of an entangled state has become the evaluation of the mean value of the Hermitian operator $\hat{W}_n$ on the state $\hat{\rho}^{\otimes n}$.
As far as we know, Eq.~\eqref{eq:witnessSVM} is a novel equation that relates a classical SVM hyperplane to a non-linear witness operator.

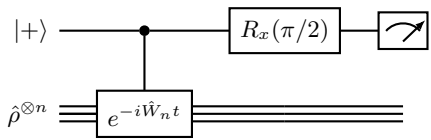
\begin{figure}
\centering
\begin{quantikz}
\lstick{$\ket{+}$} & \ctrl{1} & \gate{R_x(\pi/2)} & \meter{} \\
\lstick{$\hat{\rho}^{\otimes n}$}& [2mm] \gate{e^{-i\hat{W}_nt}}\qwbundle[alternate]{} & \qwbundle[alternate]{}& \qwbundle[alternate]{}
\end{quantikz}
\caption{Scheme for measuring the mean value of the operator $\hat{W}_n$. The control qubit is in the state $|+\rangle$. A controlled unitary gate with operator $\exp(-i\hat{W}_nt)$ is applied on the $n$ copies of the system state $\hat{\rho}$. The measurement is performed on the control qubit after a $\pi/2$ rotation $R_x$. Thus, the mean value is obtained as $\langle\hat{W}_n\rangle=\frac{p_1-p_0}{2t}$, with $p_0$ and $p_1$ being the probability that the outcome of the measurement is $0$ or $1$ respectively.\label{fig:implementation}}
\end{figure}

In order to measure the mean value of the Hermitian operator $\hat{W}_n$ we can use the procedure described in~\cite{Tacchino2020} and shown in Fig.~\ref{fig:implementation}, where a controlled unitary operator $C$-$U$, with $U=\exp(-i\hat{W}_n t)$, is applied on a system composed by the $n$ copies of the system $\hat{\rho}$. 

The control qubit is initialized in the state $|+\rangle$.
For small $t$, we can expand the operator $U\approx\mathit{1}-it\hat{W}_n+\mathcal{O}(t^2)$.
Measuring the mean value of the Pauli matrix $\langle\sigma_y\rangle$ on the control qubit corresponds to measuring $t\langle\hat{W}_n\rangle$. 
This can be achieved by transforming the $\sigma_y$ basis into the computational basis with the operator $R_x(\pi/2)=\exp(-i\sigma_x\pi/4)$, and measuring the  control qubit.
The probability of getting as outcome the state $|0\rangle$ is $p_0=1/2-t\langle \hat{W}_n\rangle$, whereas the probability of getting the outcome $|1\rangle$ is $p_1=1/2+t\langle\hat{W}_n\rangle.$ 
Calculating $p_1-p_0$ and dividing by $2t$ leads to the wanted result. 
Notice that the operator $U$ is not a local operation, and therefore, the state $\hat{\rho}$ after the protocol will not in general conserve the level of entanglement. Thus, if the goal of the circuit is to use an entangled state as input of a quantum protocol, we would need $n+1$ copies of $\hat{\rho}$.

\section{Two-qubit system}\label{sec:V}

In this section we show the results of the SVM classification for a two-qubit system. \\
In order to provide the SVM a suitable dataset, we need to uniformly sample the density matrices with respect to the Hilbert-Schmidt (HS) measure \cite{Zyczkowski2001}. Thus, we generate $4\times 4$ matrices $A$ with $\mathbb{C}$-numbers elements $A_{ij}\in\mathcal{N}(0,1)$, $\mathcal{N}(0,1)$ being the normal distribution with zero mean and unitary variance. The quantum state is obtained from the positive trace-one Hermitian matrix $\hat{\rho}=A A^\dagger/\text{Tr}[A A^\dagger]$.
Uniformly generating two-qubit quantum states accordingly to the HS measure leads to an imbalance in the classes, as the ratio between separable and entangled states is approximately 1 to 4 ($\approx 0.24$)~\cite{Slater2007}. 
We balance our dataset in order to get $50\%$ of entangled and $50\%$ of separable states. This inter-class balance allows us to have more control of the hyperparameters.
Indeed, in this case, when $\lambda_-=\lambda_+$ the SVM does not prefer any labeling on a new data point. 
Hence, we separately generate an equal number of entangled and separable states, labeling them with the PPT criterion~\cite{Horodecki1996}. 
Our dataset consists of $10^6$ samples for each class, $98\%$ of which are used in the training set, $1\%$ are used in the validation set, and $1\%$ are used in the test set. 
Each data point lives in the feature space $\mathbb{R}^{15}$, and is obtained by taking the real and imaginary part of the components of the $4\times 4$ density matrix, which is  hermitian and trace-one. 

\begin{figure}
\centering
\includegraphics[scale=0.21]{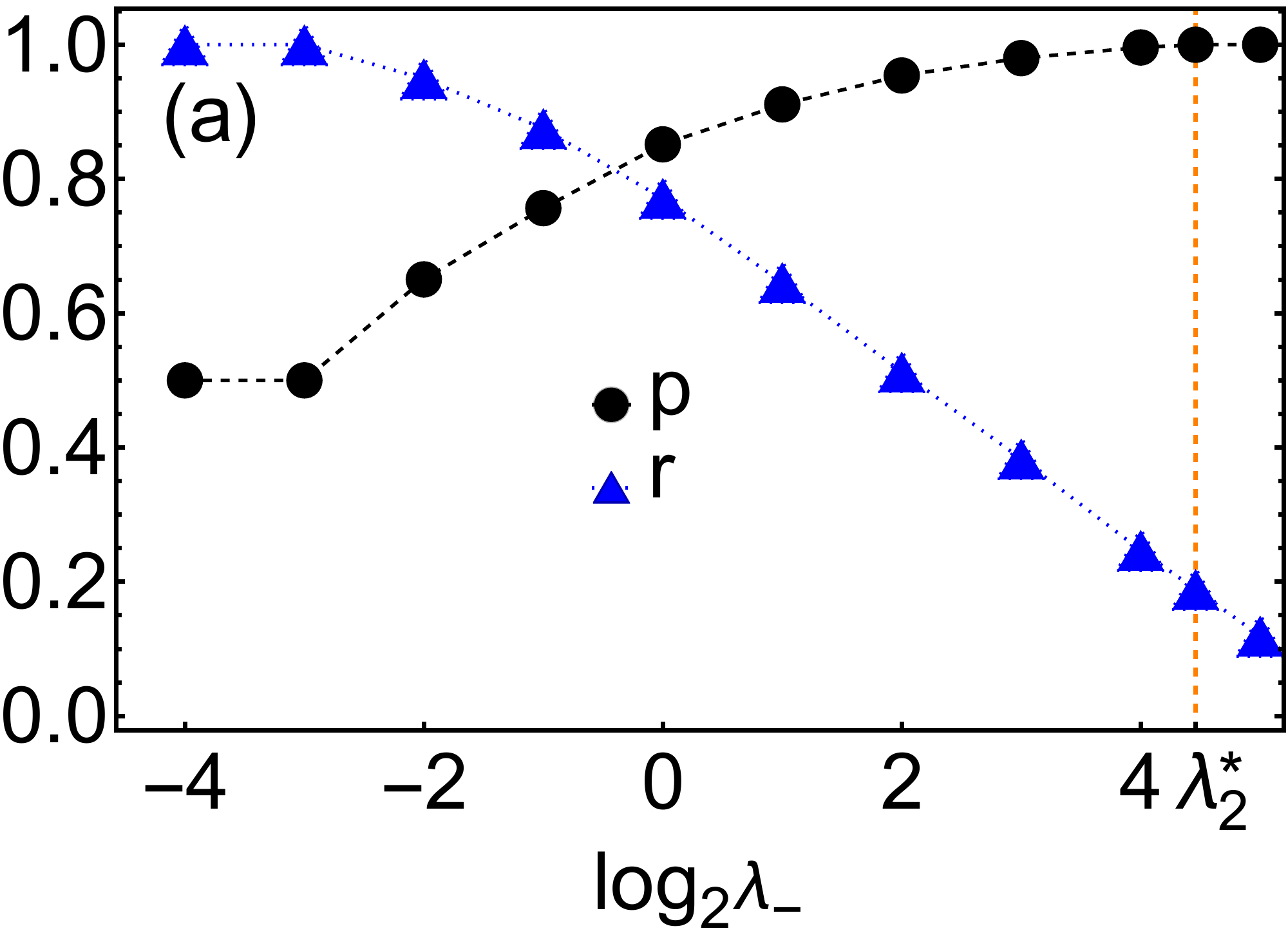}
\includegraphics[scale=0.21]{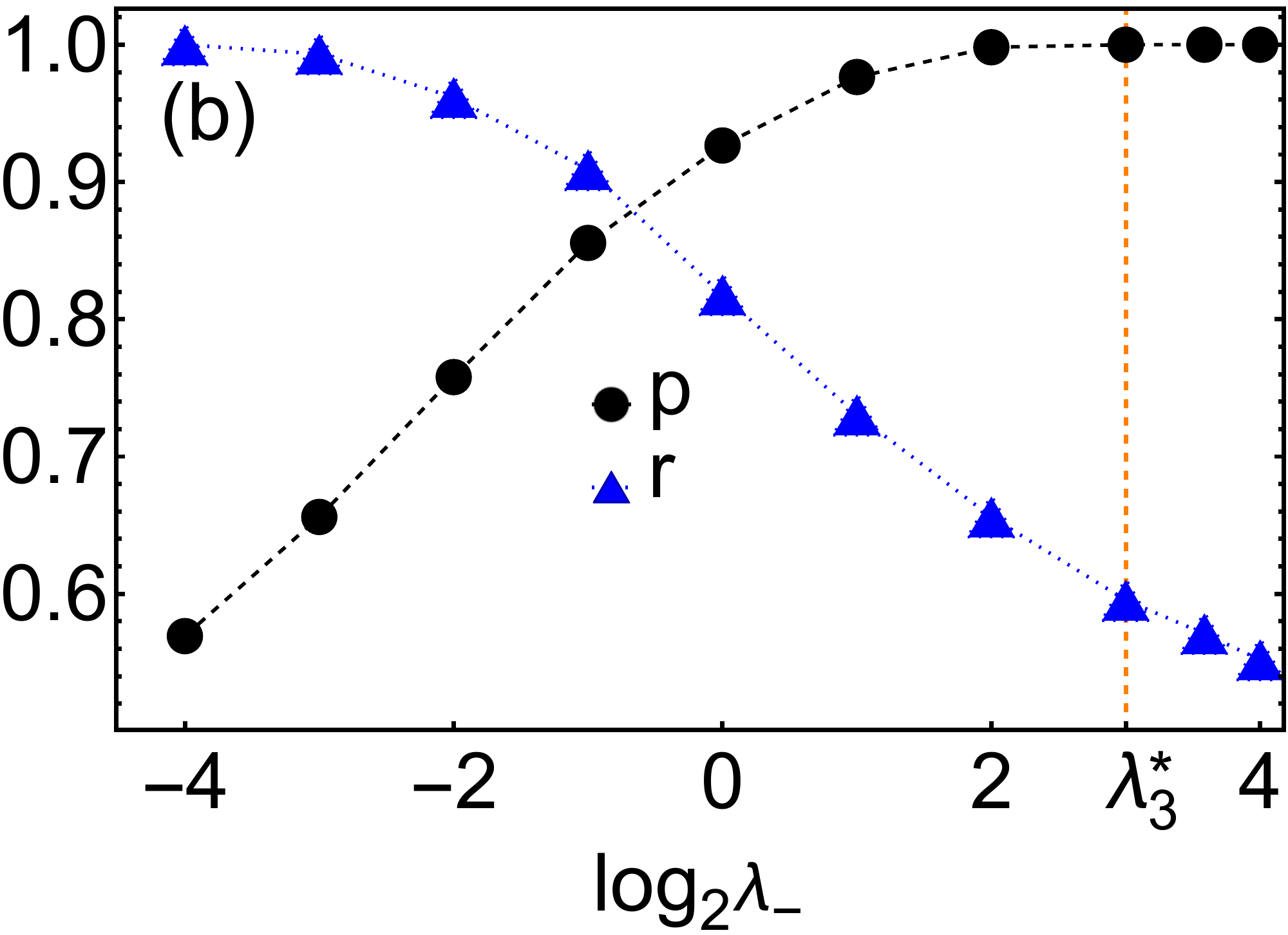}
\includegraphics[scale=0.21]{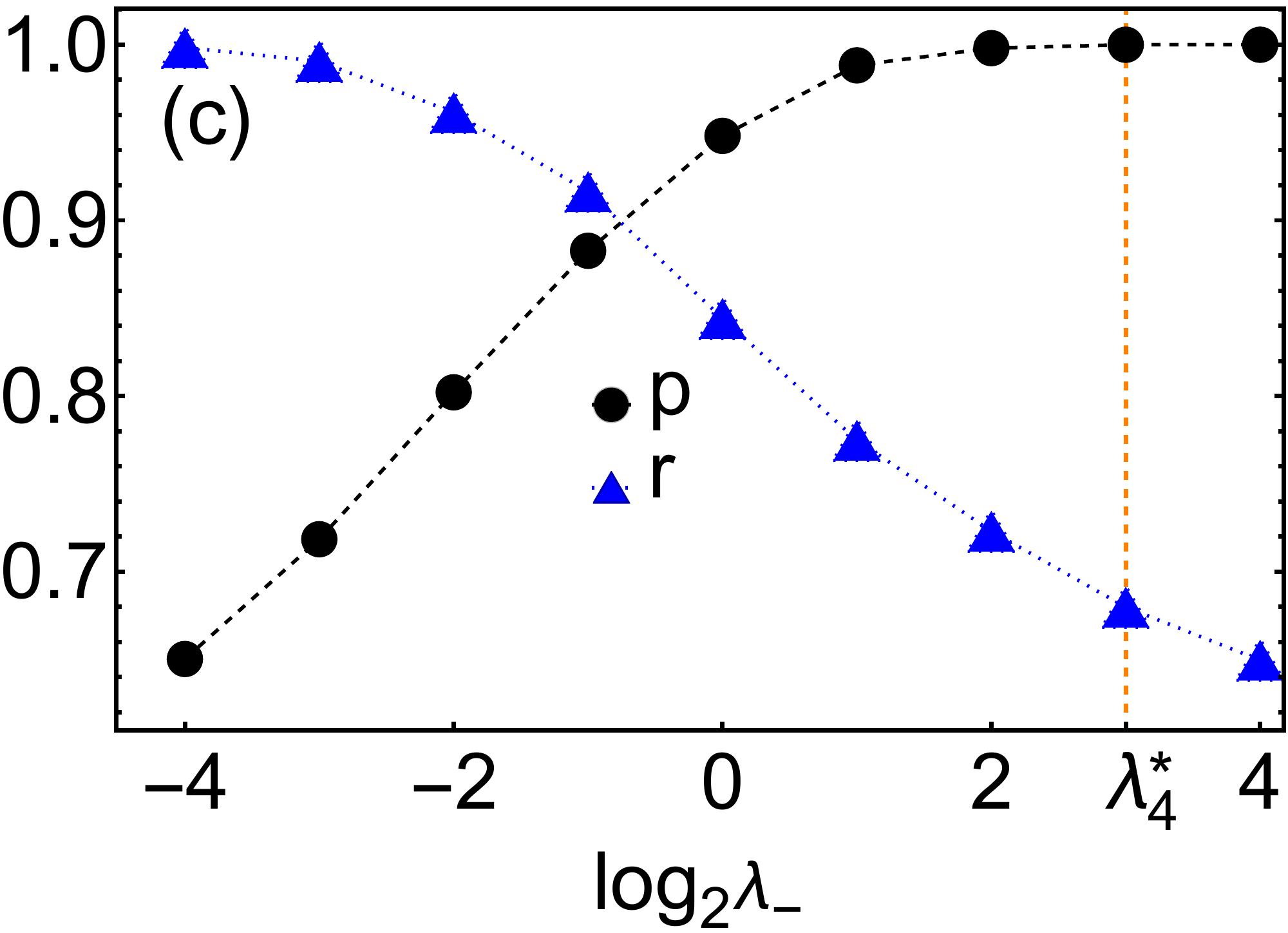}
\includegraphics[scale=0.21]{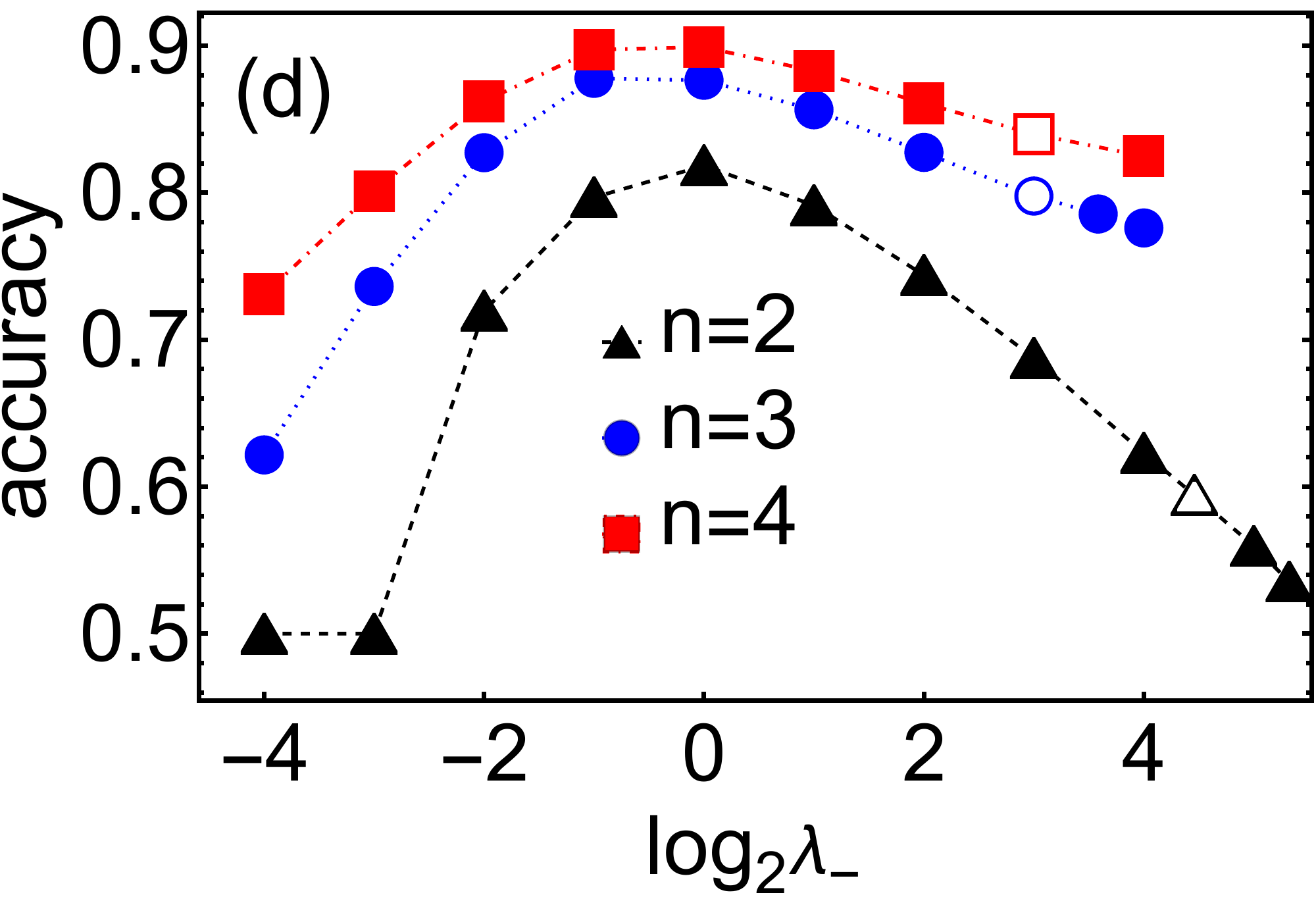}
\caption{Two qubits. Precision $p$ and recall $r$ defined in Eqs.~\eqref{eq:precision} and~\eqref{eq:recall} of the two-qubits entanglement classifier as a function of the regularization parameter $\lambda_{-}$ ($\lambda_+=1$), on a validation set of $10^4$ samples for each class in the case of polynomial kernel of degree (a) $n=2$, (b) $n=3$, and (c) $n=4$. The vertical dotted line refers to the point $\lambda_{-}=\lambda_n^*$ for each $n=2,3,4$, where $r$ is maximum with the condition $p=1$ satisfied. (d) The accuracy on the validation set for the classifier different degrees $n=2,3,4$. The empty markers correspond to $\lambda_-=\lambda_n^*$. \label{fig:2qubitspoly}}
\end{figure}

In order to fine-tune the hyperparameters, we look at the validation set. The accuracy~(\ref{eq:accuracy}), the precision and the recall~\ref{eq:recall} depend on the ratio $\lambda_{-}/\lambda_+$. Fig.~(\ref{fig:2qubitspoly}) shows the obtained precisions and recalls for different degrees $n$ of the kernel, $n=2$ (a), $n=3$ (b), $n=4$ (c), and the resulting accuracies on the validation set, calculated varying the hyperparameters $\lambda_{\pm}$. Note that using the imbalanced training set, the result would have been analogous, but shifted, and we would have got the same results using a larger value of $\lambda_+.$
For each degree of the polynomial kernel $n$, we find with a grid search the value of $\lambda_{-}/\lambda_+$ s.t. on the validation set $p=1$ and $r$ is maximum. This point, with $\lambda_{-}=\lambda_n^*$ and $\lambda_{+}=1$, is located at the orange vertical line in the plots in Figs.~\ref{fig:2qubitspoly}(a),(b) and (c). 
Fig.~\ref{fig:2qubitspoly}(d) shows the total accuracy evaluated on the validation set. We see that the classification accuracy increases for higher degree $n$.
Note that the highest accuracy is reached in correspondence of the point $\lambda_+=\lambda_{-}$, corresponding to a balanced dataset with an equal number of entangled and separable states. In fact, the value $\lambda_{-}/\lambda_+$ where we reach highest accuracy changes depending on the population of the two classes in the dataset.

\begin{figure}[h]
\centering
\includegraphics[scale=0.21]{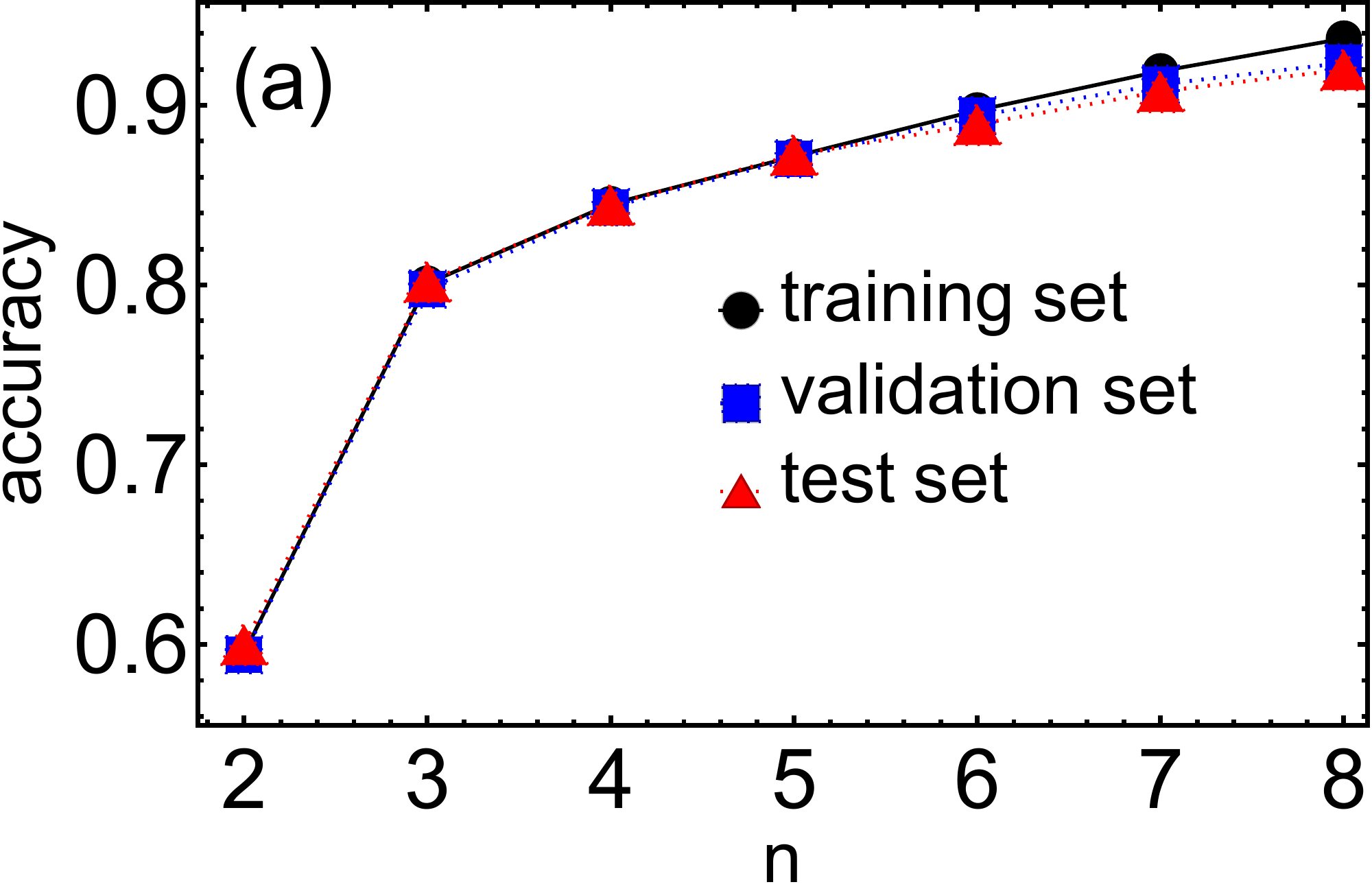}
\includegraphics[scale=0.21]{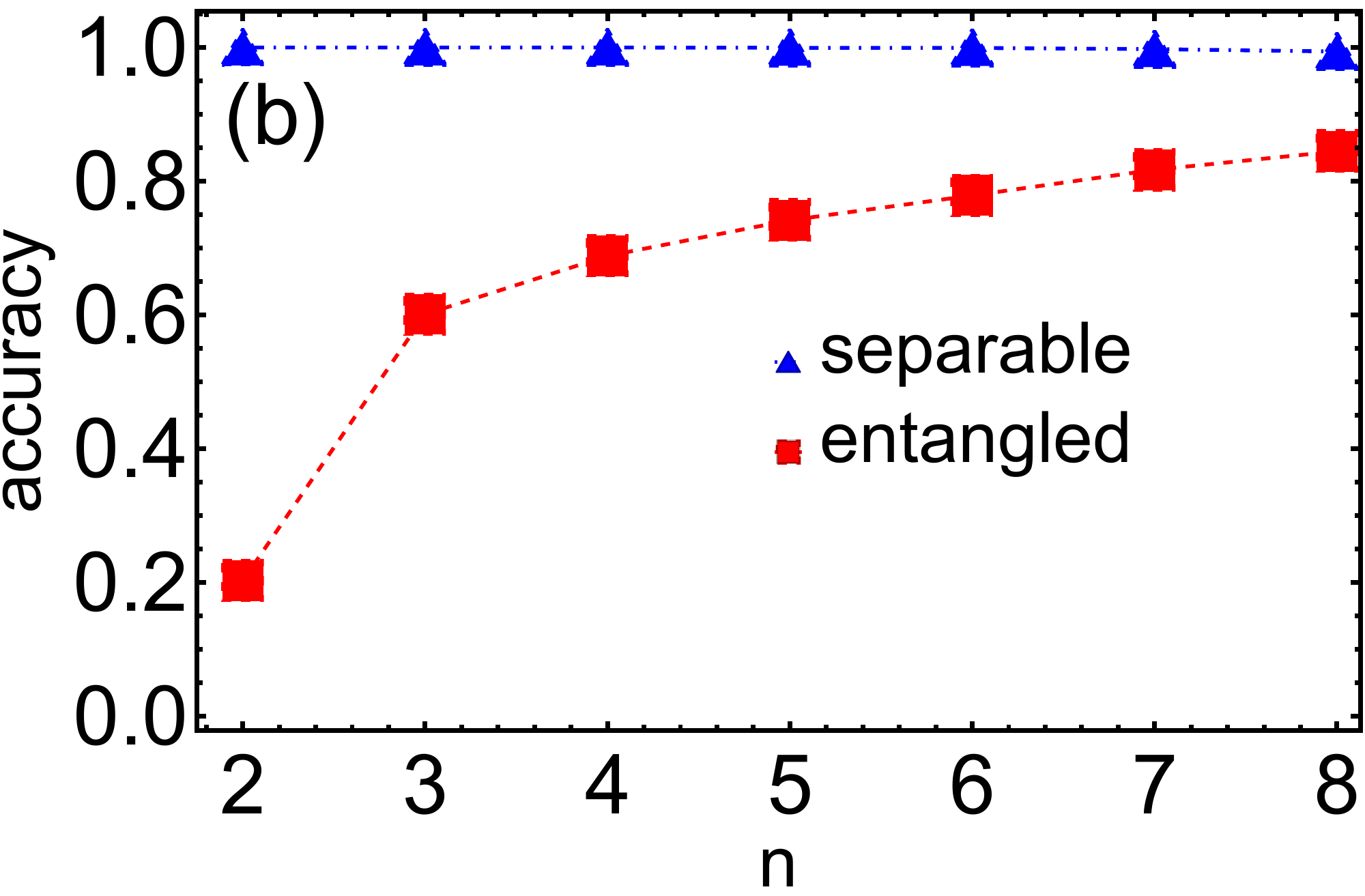}
\caption{Two qubits. (a) The accuracy obtained by the $n$ degree polynomial classifier of two-qubit states, with $n=1,2,\cdots,8$. The classifier gains accuracy with a higher $n$. The trend of the curve is the same for the three sets, showing that the classifier is able to generalize. (b) The accuracy of the classifier on the two classes.\label{fig:2qubittest}}
\end{figure}

Different degrees $n$ of the polynomial kernel lead to different performances of the algorithm. Fig.~\ref{fig:2qubittest}(a) shows the accuracy of the classifier with $\lambda_{+}=1$ and $\lambda_{-}=\lambda_n^*$ on training, validation, and test set for $n=1,2,\cdots,8$. We see that by increasing the degree $n$ we get a better accuracy in classification on the test set, reaching the value $a\approx0.91$ for the polynomial SVM with degree $8$, which indicates that
the classifier is capable of generalization. From Fig.~\ref{fig:2qubittest}(b) we see that accuracy is almost one for separable states, meaning that we selected the hyperparameters such that separable states are almost always correctly classified, while it tends to a value slightly higher than 0.8 for entangled states.

\section{Three-qubit system}\label{sec:VI}

In the same spirit as the two-qubits classification, we look for the hyperplane that minimizes the number of misclassified separable states. The goal is to be as close as possible to the border of the convex set of the separable states and to the definition of entanglement witness.
However, the three-qubit classification of entangled state has a different inherent nature, as we aim to classify our state into three different classes: the class of fully separable states, the class of biseparable states and the class of GME states. This is achieved by introducing two classifiers. We call SEP-vs-all the classifier that aims at splitting the space into fully separable states and either biseparable or GME states; this classifier defines the operator witness $\hat{W}_S$.
We call GME-vs-all the classifier that aims at splitting the GME states from the set of biseparable and fully separable states; this corresponds to the operator witness $\hat{W}_E$. 
Note that the classifiers here defined are able to distinguish only the GME states that belong to the classes that we have used in the dataset, see Table~\ref{tab:tripartiteclassification}. In fact, they are not able to classify states whose classes have not appeared in the training set. This is intrinsic in the nature of ML classifiers, that rely on known data to label new states. 

For a three-qubit system, the feature space is composed of $63$ real numbers describing the real and imaginary parts of the independent components of the $8\times 8$ hermitian and trace-one density matrix. The datasets of the two classifiers are obtained generating $2\times10^5$ states, $90\%$ of which are used in the training set, $5\%$ are used in the validation set, and $5\%$ are used in the test set.
Pure and mixed fully separable states are easily generated by construction and labeled as $-1$ for both the datasets of the two classifiers.
In the SEP-vs-all classifier, mixed fully separable states are the $50\%$ of the dataset, whereas they are $25\%$ of the dataset used for the GME-vs-all classifier.

GME states are labeled as $+1$ for both SEP-vs-all and GME-vs-all classifiers.\\
The GME states are labeled according to the three-tangle for generated pure GHZ states~\cite{Dur2000}, defined in Eq.~\eqref{eq:GHZ3}, GHZ-symmetric states~\cite{Lohmayer2006}, and statistical mixture of GHZ, W, and $\bar{W}$ states~\cite{Jung2009}. In the case of mixed X-state~\cite{Hashemi2012} we use the GME-concurrence, Eq.~\eqref{eq:GMEconcurrence}. We refer the reader to Appendix~\ref{app:I} for the definition of these states. 
Once we have generated these states, a random local unitary transformation was applied on each qubit to increase the number of data and fill the different components of the density operator.
Each category of GME states is generated with the same number of samples, and they are $25\%$ of the SEP-vs-all dataset, and $50\%$ of the GME-vs-all dataset.
The qualitative difference in the datasets of the two classifiers lies in the label of biseparable states, as given in Table~\ref{tab:tripartiteclassification}. 
For the classifier SEP-vs-all, only pure biseparable states can be considered in the dataset, as a mixture of biseparable state can also be fully separable. These states are generated and labeled as $+1$. Their category represent the $25\% $ of this dataset.
On the other hand, the dataset of classifier GME-vs-all contains mixture of biseparable states with respect to the same partition, and labeled as $-1$. This is because mixture of biseparable states with respect to different partitions can either be biseparable or fully separable, as shown in Fig~\ref{fig:setthreequbit}. They are the $25\%$ of the samples in the GME-vs-all dataset.

\begin{figure}
\includegraphics[scale=0.21]{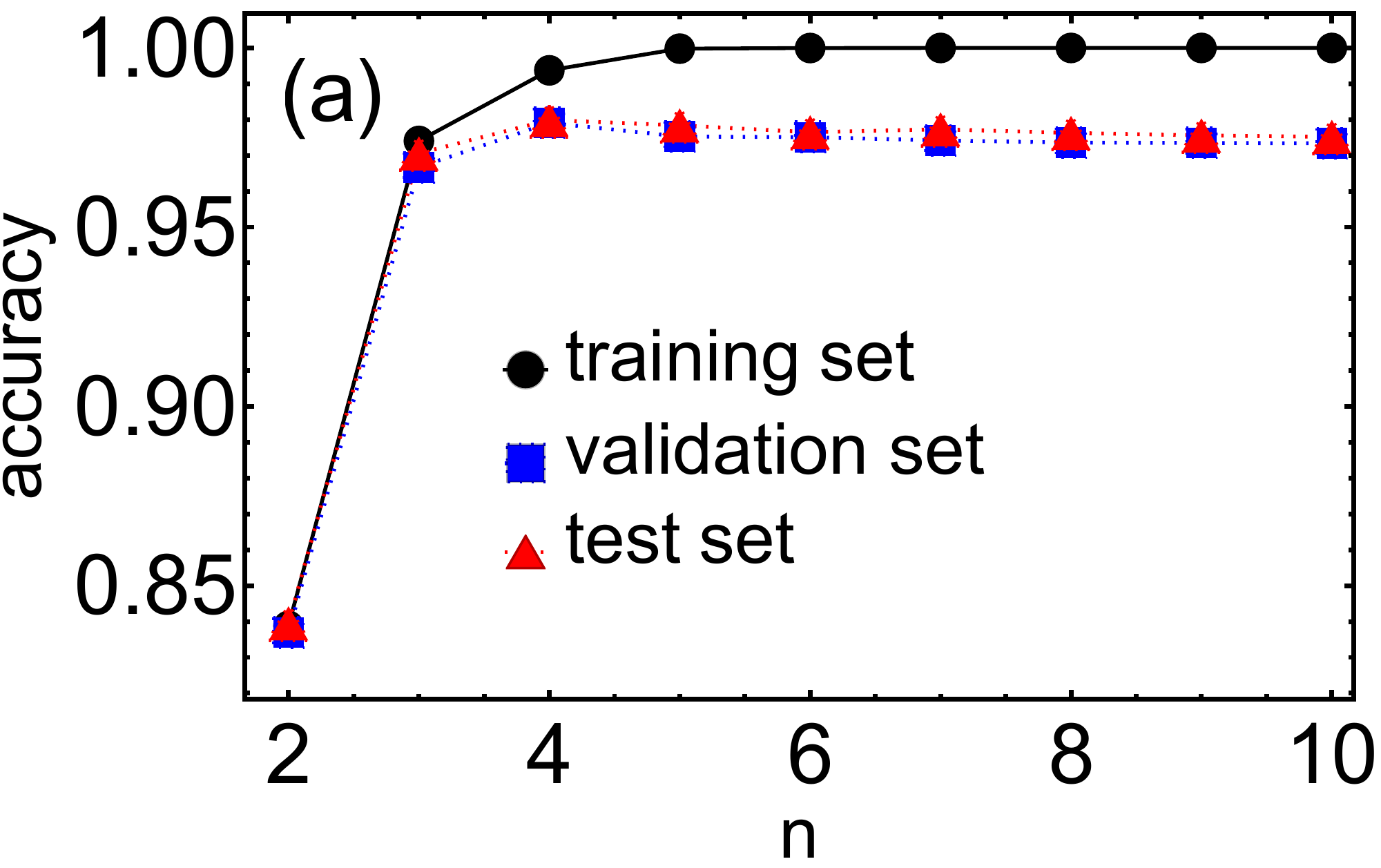}
\includegraphics[scale=0.21]{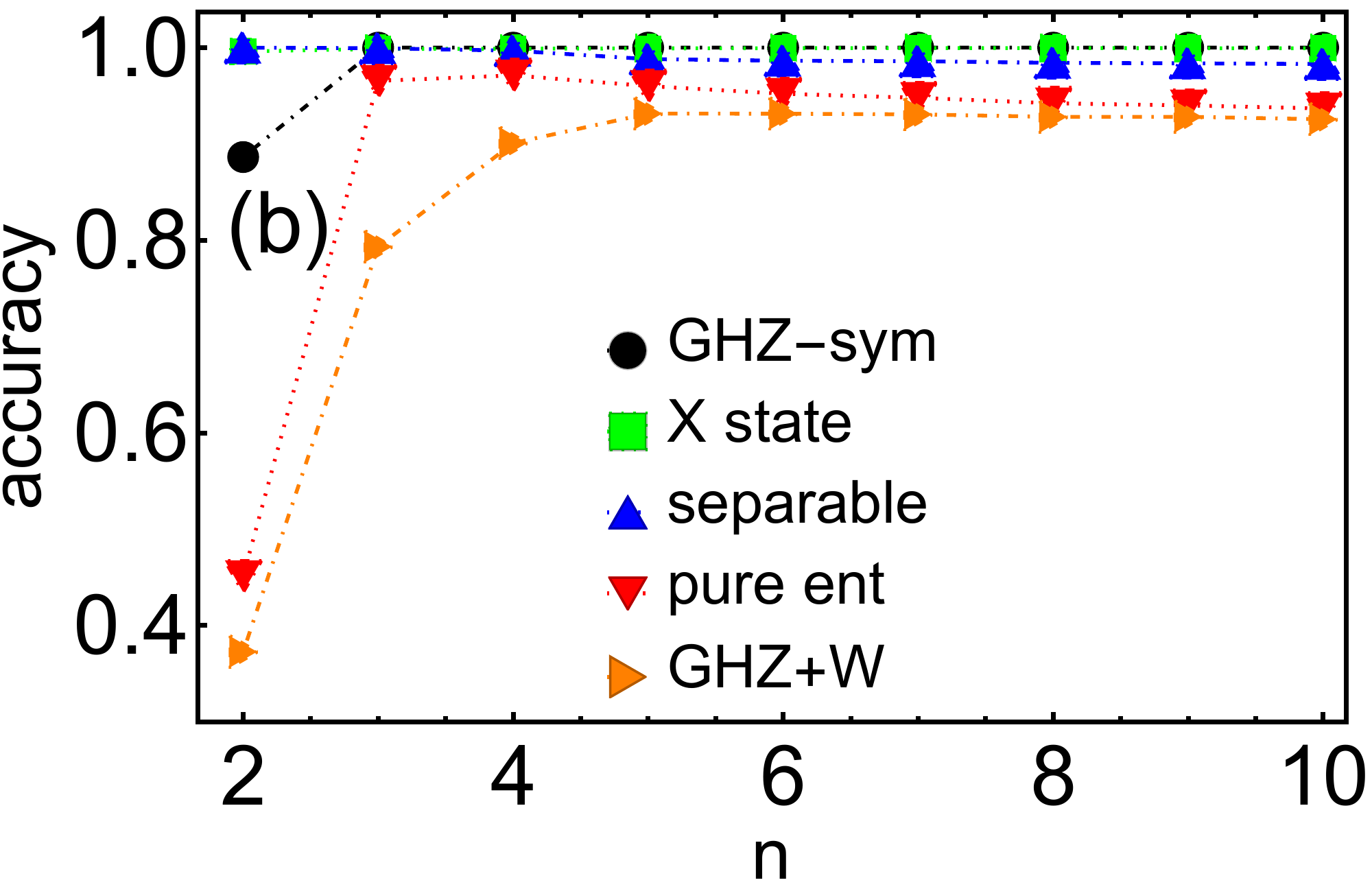}
\caption{Three qubits. The accuracy obtained by the GME-vs-all classifier. (a) The accuracy on the training, validation and test sets. (b) The accuracy for different entanglement-type states used in the training set, as a function of the kernel degree $n$. The results are shown at the fixed ratio $\lambda_{-}/\lambda_+$ obtained maximizing $r$ on the validation set with the constraint $p=1$.\label{fig:3qubitspolyGME}}
\end{figure}

\begin{figure}
\centering
\includegraphics[scale=0.21]{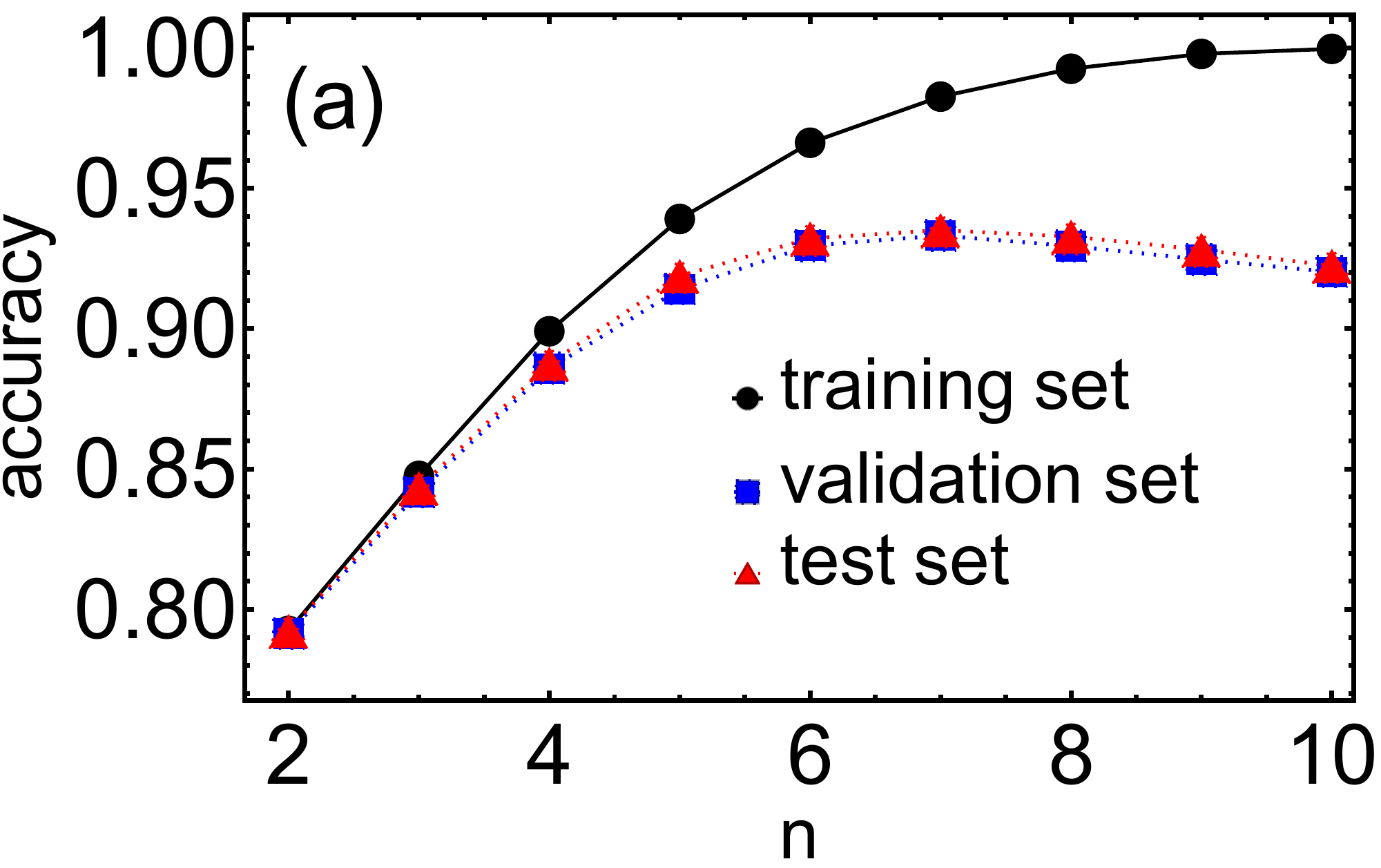}
\includegraphics[scale=0.21]{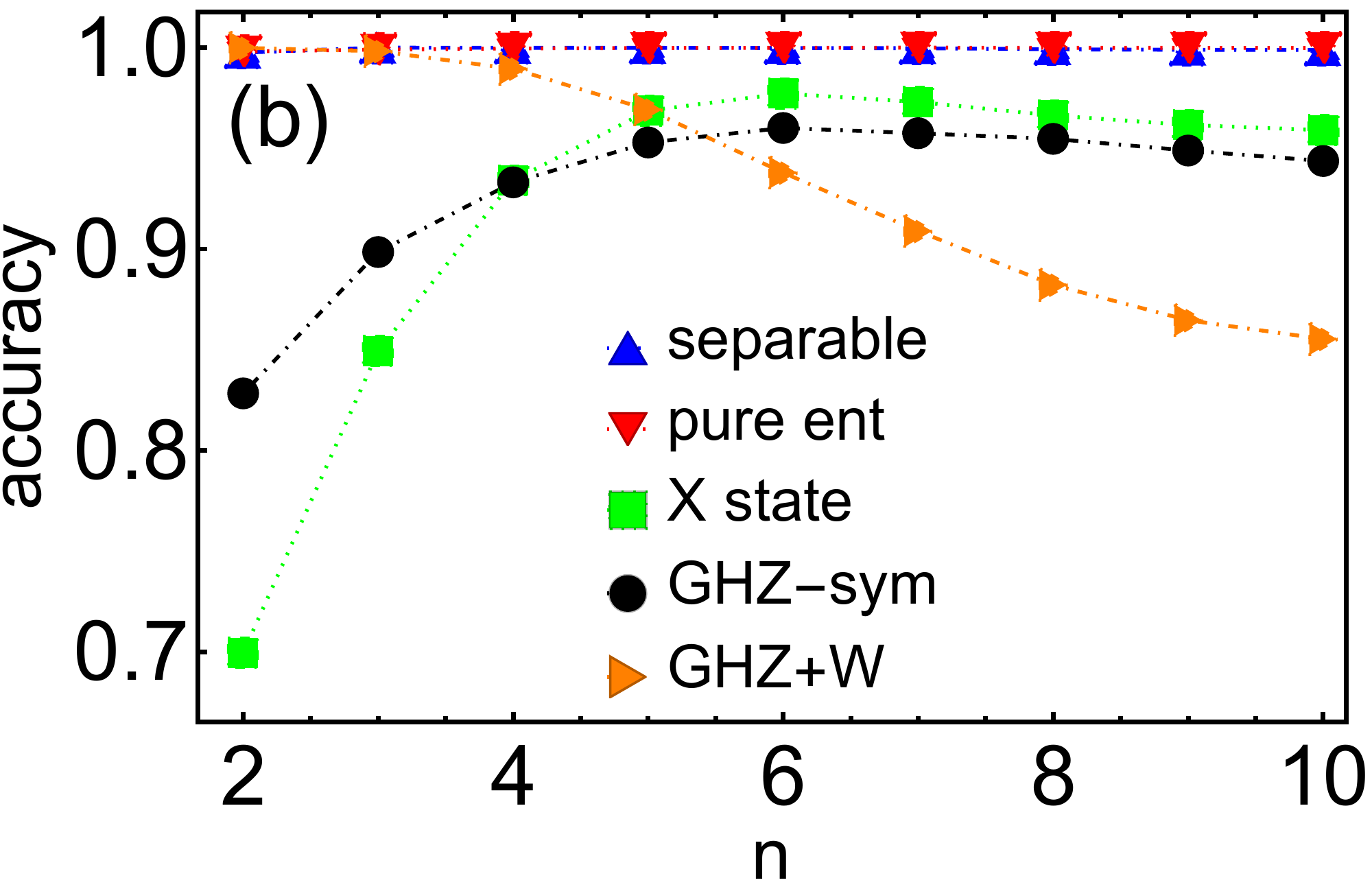}
\caption{Three qubits. The accuracy obtained by the SEP-vs-all classifier. (a) The accuracy on the training, validation and test sets.
(b) The accuracy for different entanglement-type states used in the training set, as a function of the kernel degree $n$. The results are shown at the fixed ratio $\lambda_{-}/\lambda_+$ obtained maximizing $r$ on the validation set with the constraint $p=1$.\label{fig:3qubitspolySEP}}
\end{figure}

Let us examine the accuracy of both classifiers on the different classes of states that we considered.\\
Fig.~\ref{fig:3qubitspolyGME}(a) shows the accuracy on the test set of the classifiers GME-vs-all for different values of the degree $n$, $n=2,3,\cdots,10$. Fig.~\ref{fig:3qubitspolySEP}(a) shows the accuracy obtained for the classifier SEP-vs-all on the test set. Figs.~\ref{fig:3qubitspolyGME}(b) and ~\ref{fig:3qubitspolySEP}(b) show the classification accuracy for each category of entanglement states used in the dataset on the corresponding test set. We notice that GHZ+W states are more difficult to classify for both GME-vs-all and SEP-vs-all classifiers. Instead, GHZ symmetric states are correctly labeled with an accuracy $\approx 1$ by the GME-vs-all classifier, and with accuracy $>0.9$ for $n\geq3$ by the SEP-vs-all classifier.

Contrary to what happens for the two-qubit classifier, we observe that the GME-vs-all classifier shows overfitting for $n>4$. This feature is captured by the spread between the accuracies of the training and the validation sets shown in Fig.~\ref{fig:3qubitspolyGME}(a). The same happens to the SEP-vs-all classifier for $n>6$.

\section{Conclusion}\label{sec:VII}

In this work we have investigated the possibility of using a SVM for classification of entangled states, for two-qubit and three-qubit systems.

The SVM algorithm, which is known to be suitable for binary classification of data points belonging to convex sets, learns the best hyperplane that separates the set of separable states from the set of entangled states. This construction has a strong analogy with that of an entanglement witness operator, for which the sign of the mean value calculated over the state, is related to the separability of the latter. However, witness operators are usually linear operators. 
In this paper we have shown that, exploiting the SVM algorithm and the kernel trick, we can use many copies of the state and construct non-linear witness operators, which show an increasing classification accuracy. We believe that these constructions can help us to gain knowledge about the geometry of the set of entangled states.

In the application of the SVM to the two-qubit system, we have evaluated accuracy, precision and recall for kernels of degree $n$ up to $8$, reaching an increasing accuracy.  For three-qubit systems we have trained two classifiers, named SEP-vs-all, that distinguishes fully separable states to all the others, and GME-vs-all, for detection of GME states, with kernels of degree $n$ up to $10$. The SEP-vs-all classifier shows an accuracy on the test set of $92.5\%$, whereas the GME-vs-all classifier reaches an accuracy of $98\%$,  in the best case corresponding to $n=4$. Furthermore, we have checked the accuracy among the different types of states that we have used in the dataset.

Overall, the SVM algorithm has shown a high accuracy in detecting entanglement. We have also discussed how the classifier can be trained to favor higher precision in recognition of entangled states, at the expenses of a lower global accuracy. This feature can be used in quantum computation algorithms, where entanglement plays an important role for quantum advantage. Moreover, if a quantum protocol needs a specific type of entangled state, the SVM algorithm can be trained to recognize it and use it as input. 

As entanglement classification is still an open line of research both for the theoretical and fundamental aspects, we believe machine learning algorithms can provide a useful tool of analysis.

\begin{acknowledgments}
We thank T.~Apollaro and M.~Consiglio for useful discussions.
This research is funded by the International Foundation Big Data and Artificial Intelligence for Human Development (IFAB) through the project “Quantum Computing for Applications”.
E.~E.~is partially supported by INFN through the project “QUANTUM” and the QuantERA 2020 Project “QuantHEP”.
\end{acknowledgments}
\appendix

\section{Three-qubit entangled mixed states}\label{app:I}

In this appendix we briefly describe the three-qubit GME states that we generated for the dataset of the SEP-vs-all and GME-vs-all classifiers.

A GHZ symmetric state is characterized by three fundamental properties, i) it is invariant under index permutation of the qubits, ii) it is invariant under simultaneous spin flip of the qubits, and iii) it is invariant under the rotation
\begin{equation}
U(\phi_1,\phi_2) = e^{i\phi_1\hat{\sigma}_z}\otimes e^{i\phi_2\hat{\sigma}_z}\otimes e^{-i(\phi_1+\phi_2)\hat{\sigma}_z}.
\end{equation}
This state can be written as a mixture of the two GHZ states, with density operators $\hat{\rho}_{\text{GHZ}_\pm}$, and of the maximally mixed state with density operator $\mathbb{1}/8$,
\begin{equation}
p\, \hat{\rho}_{\text{GHZ}_+}+q\, \hat{\rho}_{\text{GHZ}_-}+(1-p-q)\mathbb{1}/8\nonumber.
\end{equation}
When $q=0$, we have the so-called Werner state~\cite{Werner1989}, with a GHZ state combined with the maximally mixed state. 
In Refs. \cite{Eltschka2014,Eltschka2008,Eltschka2012}, Eltschka et al. gave a complete classification of GHZ symmetric states in terms of the relation between the coefficients $p$ and $q$.

The X-state owes its name to the shape of the density matrix, which has non null values on the diagonal and anti-diagonal elements. Thus the X-state density operator can be written as the sum $\hat{D}+\hat{A}$, with $\hat{D}=\text{diag}(a_1,a_2,\dots,a_n,b_n,\dots,b_1)$ the $2n\times 2n$ diagonal matrix, and $\hat{A}=\text{anti-diag}(z_1,\dots,z_n,z_n^*,\dots,z_1^*)$ being an anti-diagonal matrix. The conditions $Tr[\hat{D}]=1$ and $\hat{D}\geq\hat{A}$ ensure that the operator $\hat{D}+\hat{A}$ describes a physical state. The GME-concurrence of the X-state can be analytically calculated~\cite{Hashemi2012}, and when its value is positive, the X-state belongs to the GME class. Note that the GHZ symmetric states are particular X-states with $z_i=0$ for $i= 2,\dots,n$.

Another class that we have used to train our classifier is the statistical mixture of GHZ, W, and the flipped W state $\bar{\text{W}} = \frac{1}{\sqrt{3}}(|011\rangle+|101\rangle+|110\rangle)$, that we have labeled as GHZ+W, with density operator 
\begin{equation}
p\, \hat{\rho}_{\text{GHZ}}+q\hat{\rho}_{\text{W}}+(1-p-q)\, \hat{\rho}_{\bar{\text{W}}}.\nonumber
\end{equation}
The analytic calculation of the three-tangle measure for those states has been given in \cite{Eltschka2008,Lohmayer2006,Jung2009}.

\bibliographystyle{unsrt}
\bibliography{SVMentanglement}

\end{document}